\documentstyle[array,english,prl,aps,multicol]{revtex}

\newcommand{\ket}[1]{|{#1} \rangle}
\newcommand{\bra}[1]{{\langle {#1}|}}
\newcommand{\braket}[2]{\langle {#1} | {#2} \rangle}

\newcommand{\Ezz}[0]{E_{0,0}}
\newcommand{\Ezu}[0]{E_{0,1}}
\newcommand{\Euz}[0]{E_{1,0}}
\newcommand{\Euu}[0]{E_{1,1}}
\newcommand{\Eij}[0]{E_{i,j}}
\newcommand{\Id}[0]{{\openone}}

\begin{document}

\title{Security of Quantum Key Distribution Against All Collective
Attacks}

\author{Eli Biham$^{(1)}$, Michel Boyer$^{(2)}$, Gilles Brassard$^{(2)}$, Jeroen van de Graaf$^{(2)}$ and Tal Mor$^{(2)}$ 
 }
\address{ (1) Computer Science Department, Technion, Haifa 32000, Israel; (2) DIRO, Universit\'e de Montr\'eal, Montr\'eal, Canada; 
}

\date{\today}

\maketitle

\begin{abstract}
Security of quantum key distribution against sophisticated attacks
is among the most important issues in quantum information theory.
In this work we prove security against a very important class of attacks
called {\em collective attacks} (under a compatible noise model)
which use quantum memories and 
gates, and which are directed against the final key.
Although attacks stronger than 
the collective attacks can exist
in principle, no explicit example was found and it is conjectured that
security against collective attacks implies also security against any
attack.
\end{abstract}

\begin{multicols}{2}

\paragraph{Introduction}


Quantum cryptography is one of the most surprising 
consequences of processing information using 
quantum two-state systems (qubits) instead of 
classical bits. 
Quantum key distribution was invented 14 years ago~\cite{BB84}, 
to provide a new type of solution to
one of the most important cryptographic problems, the transmission
of secret messages. 

For many years physicists and computer scientists
have been trying to prove the security of various
quantum key distribution schemes. 
Many particular ``simple''  cases where analyzed,
such as the {\em intercept-resend attacks} and 
the {\em individual particle attacks}, for which there is a clear intuition
that classical privacy amplification provides the 
desired security,
but no explicit bound on the information available
to an eavesdropper has been proven.
The security in case of the general {\em joint attacks} which are using
quantum gates and quantum memory and are 
directed against the {\em final key} was also considered in several 
works (see~\cite{BM97b,Mayers} and references there in). 
In this paper we complete the work started in~\cite{BMS96,BM97a,BM97b} 
to conclude 
that the four-state scheme~\cite{BB84} for 
quantum key distribution is secure against any {\em collective attack}
(an important subclass of the joint attacks) under a compatible error model.

In the four-state scheme Alice sends to Bob a classical string of length
$n''$ using a quantum channel, by sending qubits;
she sends either $|0\rangle_z$ or 
$|0\rangle_x = (|0\rangle_z + |1\rangle_z)/\sqrt2 $ to encode a bit value $0$,
or she sends either $|1\rangle_z$ or 
$|1\rangle_x = (|0\rangle_z - |1\rangle_z)/\sqrt2 $ to encode a bit value $1$.
Alice and Bob are also connected by a classical channel 
which is insecure but unjammable.
At a later stage Alice tells Bob (classically), regarding each qubit, 
whether she used the $z$ basis 
or the $x$ basis. If Bob has used the same basis for his measurement,
they keep the bit (which is supposed to be the same as Alice's),
so they are left with $n'$ similar bits.
Alice and Bob now estimate the error rate using some ($n_{test}$) test bits. 
If the estimated error-rate $p_{test}$ is less than some pre-agreed threshold
$p_{allowed}$, then the test 
succeeds and Alice and Bob obtain a final key from the remaining $n$-bit string
(where $n=n'-n_{test}$),
by performing error correction and privacy amplification.
They choose parities of $k$ substrings for error-correction and 
parities of $m$ substrings
for privacy amplification. The parity of each of the $k$ 
substrings is announced
in order to correct the string,
and the parities of the $m$ substrings are kept secret, and  
used as the final key.
We consider $m=1$ in the following and leave the general case 
to a review paper.

In the most general (so called ``joint'') attack, 
Eve can do whatever she likes (the most general
unitary transformation using an ancila) to the qubits, and delay all her 
measurements till receiving all classical data. 
We restrict ourself to ``collective'' attacks~\cite{BM97a} 
where each qubit is attached to 
a separate probe (unentangled to the other probes), and the measurement 
is delayed, and is performed collectively on all probes, after all 
classical data is obtained.
There are good reasons~\cite{BM97a,BM97b} to believe that collective attacks 
are the strongest joint attacks (when $n$ is large).
Furthermore, no particular joint attack was shown 
to be stronger than collective attacks.
In a collective attack, 
after Alice sends $n''$ qubits to Bob, each is attached to a separate 
probe by Eve. Then, the global state of the Eve-Bob system is
$ \rho_1 \otimes \ldots \otimes \rho_{n''} $  where each 
$\rho_i$ is a density operator on the space ${\cal H}^{E_i} \otimes
{\cal H}^{B_i}$ where the spaces ${\cal H}^{E_i}$ and ${\cal H}^{B_i}$ belong
respectively to Eve and Bob. 

\paragraph{Bounds on information}

We shall first fix some notations from information theory.
Let $B$ and $X$ be random variables (describing the
input and output of a channel).
When the context is clear we write 
$p(b)$ for $p(B \! = \! b)$ and  
$p(x)$ for $p(X \! = \! x)$.
The joint probability $p(x,b)$ satisfies
$p(x) = \sum_{b \in B} p(x,b)$ and 
$p(b) = \sum_{x \in X} p(x,b)$.
The conditional 
probability is denoted by 
$p_b(x) \equiv p( X = \! x  | B \! = \! b  ) $ and
$p_x(b) \equiv p( B = \! b  | X \! = \! x  ) $.
It satisfies the Bayes formula    
$p_b(x) p(b) = p(x,b) = p_x(b) p(x)$. 
The mutual information between the input and the output 
probability distributions,  
$I(X;B)= - \sum_{b \in B} p(b) log_{{}_2} p(b) + 
\sum_{x \in X} p(x) \sum_{b \in B} p_x(b) log_{{}_2} p_x(b) $,
tells us the increase of knowledge about the input, if the output
becomes known to us.

For a binary input $B$ with equal input probabilities
$I(B;X) 
= 
\sum_{x \in X} p(x) I_2(p_x(0))$,
where $I_2(p) = 1 + p \log_{{}_2} p + (1-p) \log_{{}_2} (1-p)$.
Distinguishing the input when the output
is given, is then 
equivalent to distinguishing the two probability distributions
$p_0(x)$ and $p_1(x)$.
All the probability distributions in the expression of the mutual information
can be calculated from $p_0(x)$, $p_1(x)$, so we can define 
another function SD, Shannon Distinguishability,
$SD(p_0(x),p_1(x)) \equiv I(B;X)$ (restricted to binary input with equal 
probabilities).

Suppose we are given a state (density matrix) $\rho$.
The most general measurement giving a
result $x$ in some set $X$ of possible outputs is given
by a POVM~\cite{Peres} 
indexed by $X$, i.e.\ a family ${\cal E} = (E_x)_{x \in X}$ of 
Hermitian operators $E_x$ with non negative eigenvalues
such that $\sum_{x \in X}E_x=\Id$. The probability
of occurrence of $x$ given the state $\rho$ is then equal to 
$p^{\cal E}(x) = Tr(\rho E_x)$. 
Given two equally likely states $\rho_0$ and $\rho_1$, 
and a measurement procedure 
${\cal E}$, let
$p_b^{\cal E}(x) = Tr(\rho_b E_x)$;
for any given ${\cal E}$ let 
$SD^{\cal E}(\rho_0, \rho_1) \equiv 
SD(p^{\cal E}_0(x), p^{\cal E}_1(x))$.
The maximum information 
we can get regarding the state we are facing 
is given by the optimal Shannon Distinguishability 
$
SD(\rho_0, \rho_1) \equiv \sup_{\cal E}[ SD^{\cal E}(\rho_0, \rho_1) ]
$
where the supremum is taken over all POVM's on all possible sets X.

Unfortunately, there is no known analytic formula giving 
optimal mutual information.
In what follows, we will present two bounds which are
simple to state and to derive, and which will be found very useful.
\newline
{\bf {\em Theorem 1}.}~---
If $\widetilde{\rho_0}$ and $\widetilde{\rho_1}$ are two density matrices
defined on some space ${\cal H}_1 \otimes {\cal H}_2$ and
$\rho_i=Tr_2(\widetilde{\rho_i})$ are the density matrices on ${\cal H}_1$ 
obtained by tracing-out ${\cal H}_2$, then,
\begin{equation}
\label{TrMI}
SD(\rho_0, \rho_1) = SD(Tr_2(\widetilde{\rho_0}), Tr_2(\widetilde{\rho_1})) 
\leq
SD(\widetilde{\rho_0}, \widetilde{\rho_1})
\ .
\end{equation}
{\bf {\em Proof}.}~---
If ${\cal E} = (E_x)_{x \in X}$ is a POVM on ${\cal H}_1$ then
${\cal E} \otimes \Id_{{\cal H}_2} \equiv 
(E_x \otimes \Id_{{\cal H}_2})_{x \in X}$
is a POVM on ${\cal H}_1 \otimes {\cal H}_2$ 
and $Tr_1(Tr_2(\widetilde{\rho_i}) E_x) = Tr(\widetilde{\rho_i} (E_x \otimes \Id_{{\cal H}_2}))$.
Consequently
$SD^{\cal E}(Tr_2(\widetilde{\rho_0}), Tr_2(\widetilde{\rho_1})) = 
SD^{{\cal E} \otimes \Id_{{\cal H}_2}}(\widetilde{\rho_0}, \widetilde{\rho_1})$.
By definition of the optimization process 
$\sup_{\cal E} [SD^{{\cal E} \otimes \Id_{{\cal H}_2}}(\widetilde{\rho_0}, 
\widetilde{\rho_1})] \le
\sup_{\widetilde{\cal E}} [ SD^{\widetilde{\cal E}}(\widetilde{\rho_0}, \widetilde{\rho_1})]$,
and thus, 
$ SD(Tr_2(\widetilde{\rho_0}), Tr_2(\widetilde{\rho_1})) 
\leq SD(\widetilde{\rho_0}, \widetilde{\rho_1})$.
The $\widetilde{\rho_b}$ will be called a {\em lift-up} of 
$\rho_p$, and it is known as {\em purification} if it is a pure state.

This theorem (proven independently in~\cite{Fuchs})
actually states that tracing out cannot increase information.
It provides a useful upper bound on the mutual information that can be
obtained about mixed states $\rho_i$, if we can find
appropriate states $\widetilde{\rho_i}$. 
A similar idea which says that mixing cannot improve information was used 
in~\cite{BM97b} to obtain a more limited security result.

For any two density matrices $\rho_0$ and $\rho_1$  we can define 
$ Tr | \rho_0 - \rho_1 | $ 
the trace-norm of the Hermitian operator $\rho_0 - \rho_1$.
In our context where we only
consider Hermitian matrices, $Tr |A|$ is nothing but the sum of the absolute
values of the eigenvalues of $A$.
It is relatively easy to calculate the trace-norm of 
$\rho_0 - \rho_1$. Therefore, the following upper bound is very important.
\newline
{\bf {\em Theorem 2}.}~---
For any two density matrices $\rho_0$ and $\rho_1$,
\begin{equation}
\label{Jeroen}
SD(\rho_0,\rho_1) \leq \frac{1}{2} Tr | \rho_0 - \rho_1 | \ .
\end{equation}
{\bf {\em Proof}.}~---
In order to prove this equation (see also \cite{Jeroen-Fuchs}) 
let us first fix some measurement procedure
${\cal E} = (E_x)_{x \in X}$. 
Then $SD^{\cal E}(\rho_0, \rho_1) =
I(B;X) = 
\sum_{x \in X} p(x) I_2(p_x(0))$,
where (from the Bayes formula)  
$
p_x(0) = p(B \! = \! 0) p( X \! = \! x | B \! = \! 0) / p(x) = 
(1/2) p_0^{\cal E}(x)/p(x).
$
Knowing that $I_2(r) \leq | 2r - 1 | $ for $0 \leq r \leq 1$ 
we conclude that
$SD^{\cal E}(\rho_0, \rho_1) \le 
\sum_{x \in X} p(x) |2 p_x(0) -1| $. 
Assigning $p_x(0)$ into the last expression  
[and using 
$ p(x) = (p_0^{\cal E}(x) + p_1^{\cal E}(x))/2 $
in the following equality], we obtain 
$SD^{\cal E}(\rho_0, \rho_1) \le 
\sum_{x \in X} p(x) |2 [p^{\cal E}_0(x)/2p(x)] -1| = 
\frac{1}{2} \sum_{x \in X} | p^{\cal E}_0(x) - p^{\cal E}_1(x) | $.
Now, since $\rho_0 - \rho_1$ is Hermitian, it can be diagonalized
and consequently written in the form 
$\rho_0 - \rho_1 = \sum \lambda_j \ket{j}\bra{j}$ where $\ket{j}$ is an
orthonormal basis and $Tr|\rho_0 - \rho_1| = \sum | \lambda_j |$.
Clearly $Tr (\ket{j}\bra{j} E_x) = \bra{j} E_x \ket{j}$ 
and so
$p_0^{\cal E}(x) - p_1^{\cal E}(x) = 
Tr( (\rho_0 - \rho_1) E_x) =
\sum_j \lambda_j \bra{j} E_x \ket{j}$.
Using the last expression for $SD$ and using 
$\bra{j} E_x \ket{j} \geq 0$ (since $E_x$ is positive
definite), we can now deduce
$
SD^{\cal E}(\rho_0, \rho_1) \leq 
   \frac{1}{2} \sum_j | \lambda_j | \sum_{x \in X} \bra{j} E_x \ket{j} 
    = \frac{1}{2} Tr |\rho_0 - \rho_1 |.
$
Since ${\cal E}$ is arbitrary, we choose the one which optimizes SD and
this concludes the proof.

\paragraph{Error versus information}

Let us assume that Eve is powerful enough to control the natural noise.
Without loss of generality, we assume that
Eve's probes
are in some arbitrary but fixed initial (tensor product) 
pure state, and that 
each probe is
in a state
$\ket{E}$.
In the {\em collective attack},
the state $\ket{E} \otimes \ket{b}$ is subjected to Eve's 
unitary transformation 
${\cal U}$ that changes  
the state $\ket{b}$ sent by Alice to
the final global state 
\begin{mathletters}
\begin{eqnarray}
\label{TransUa}
\ket{0}_z & \mapsto & \ket{\Ezz^z} \ket{0}_z + \ket{\Ezu^z} \ket{1}_z 
\equiv  \ket{\phi_0^z} \\
\label{TransUb}
\ket{1}_z & \mapsto & \ket{\Euz^z} \ket{0}_z + \ket{\Euu^z} \ket{1}_z  
\equiv \ket{\phi_1^z}
\end{eqnarray}
\end{mathletters}
where the $\ket{\Eij^z}$ are Eve's non normalized states. 
Implicitly, this description corresponds to restricting 
natural noise to follow the spirit of the collective attacks.
It is reasonable to suspect that more general noise models would not 
be to Eve's advantage.

Bob's error probability in the $z$ basis
[measuring $\ket{0}_z$ if $\ket{1}_z$ was sent etc.] is 
$p_e^z = (1/2) 
[\braket{\Ezu^z}{\Ezu^z} + \braket{\Euz^z}{\Euz^z}]$.
Alice can also use the alternate basis $x$,
and then the transformation ${\cal U}$
can also be expressed in the $x$ basis (replacing
everywhere $z$ by $x$) to yield  
$p_e^x = (1/2) 
[\braket{\Ezu^x}{\Ezu^x} + \braket{\Euz^x}{\Euz^x}]$.
Since Alice uses both bases with the same probability, Bob's overall
probability of error is $p_e = \frac{1}{2}(p_e^x + p_e^z)$ and so 
$p_e^x \leq 2 p_e$ and $p_e^z \leq 2 p_e$. 
Due to linearity of the transformation ${\cal U}$ we obtain
$
\ket{\Ezu^x} = \frac{1}{2}[(\ket{\Ezz^z} - \ket{\Euu^z}) + (\ket{\Euz^z} - 
   \ket{\Ezu^z})]
$
and
$
\ket{\Euz^x} = \frac{1}{2}[(\ket{\Ezz^z} - \ket{\Euu^z}) - (\ket{\Euz^z} - 
  \ket{\Ezu^z})]$.
If we expand $p_e^x$ in terms of
the vectors in the $z$ basis 
we get $p_e^x = (1/4)[\langle E_{0,0}^z - E_{1,1}^z | 
                               E_{0,0}^z - E_{1,1}^z \rangle +
                       \langle E_{1,0}^z - E_{0,1}^z | 
                               E_{1,0}^z - E_{0,1}^z \rangle ]$.
Since ${\cal U}$ preserves inner products, the states 
$\ket{\phi_0^z}$ 
and 
$\ket{\phi_1^z}$ 
have norm 1. Therefore,  
$\braket{\Ezz^z}{\Ezz^z} + \braket{\Ezu^z}{\Ezu^z}=1$ and 
$\braket{\Euz^z}{\Euz^z} + \braket{\Euu^z}{\Euu^z}=1$, which we use to get
\begin{equation}
  \label{BndPe}
  p_e^x = \frac{1}{2}[ 1 - Re\{ \braket{\Ezz^z}{\Euu^z} + 
\braket{\Euz^z}{\Ezu^z}
  \} ]  
\ .
\end{equation}

Eve's view is 
obtained by tracing-out Bob from the states $\phi_b^z$
(if the $z$ basis was used):
$\rho_0^z(E) = \ket{\Ezz^z}\bra{\Ezz^z} + \ket{\Ezu^z}\bra{\Ezu^z}$ and 
$\rho_1^z(E) = \ket{\Euz^z}\bra{\Euz^z} + \ket{\Euu^z}\bra{\Euu^z}$.
Many other pure states (purifications)
also yield the same reduced density matrices for Eve.
In particular 
\begin{mathletters}
\begin{eqnarray}
\label{Psis}
\ket{\psi_0^z} &=& \ket{\Ezz^z}\ket{0}_z+\ket{\Ezu^z}\ket{1}_z \\
\ket{\psi_1^z} &=& \ket{\Euu^z}\ket{0}_z+\ket{\Euz^z}\ket{1}_z
\end{eqnarray}
\end{mathletters} 
will prove useful 
since the angle between them is zero if there is no disturbance.
While these states have only virtual existence, they will be used 
in Theorem 1 to yield the desired bound, since Eve's states
are the trace-out of these pure states.
They live in some Hilbert space 
${\cal H}^E \otimes {\cal H}^2$, with ${\cal H}^2$ 
two-dimensional Hilbert space. They are
normalized, and consequently
$|\braket{\psi_0^z}{\psi_1^z}|
= cos(2\alpha_z)$ for some angle $0 \leq \alpha_z
\leq \pi/4$. 
Moreover, there is some phase angle $\theta$ such that
$e^{i\theta} \braket{\psi_0^z}{\psi_1^z} =
|\braket{\psi_0^z}{\psi_1^z}|$. Let $\ket{\Psi_0^z} = \ket{\psi_0^z}$ 
and $\ket{\Psi_1^z} = e^{i\theta}\ket{\psi_1^z}$. 
One can now find two (normalized) orthogonal states $\ket{0^z_{\cal H}}$
and $\ket{1^z_{\cal H}}$ (spanning a two dimensional subspace ${\cal H}$ of 
${\cal H}^E \otimes {\cal H}^2$) such that 
$\ket{\Psi_0^z} = 
  \cos(\alpha_z)\ket{0^z_{\cal H}} + \sin(\alpha_z)\ket{1^z_{\cal H}}$
and
$\ket{\Psi_1^z} = 
  \cos(\alpha_z)\ket{0^z_{\cal H}} - \sin(\alpha_z)\ket{1^z_{\cal H}}$.
{}From 
$1 - 2 \sin^2(\alpha_z) =
\braket{\Psi_0^z}{\Psi_1^z} =
|\braket{\psi_0^z}{\psi_1^z}| =
|\braket{\Ezz^z}{\Euu^z} + \braket{\Euz^z}{\Ezu^z} |$
we deduce that
$Re\{ \braket{\Ezz^z}{\Euu^z} + \braket{\Euz^z}{\Ezu^z} \} \le 
1 - 2 \sin^2(\alpha_z)$.
Using (\ref{BndPe}) we get that Eve's state is a partial trace of one of
the pure states $\ket{\Psi_b^z}$ with angle satisfying 
$ \sin(\alpha_z) \leq (p_e^x)^{1/2} $.

Everything that has been said about $|\psi_b^z\rangle$ and $p_e^x$ 
holds by symmetry for replacing the bases,
yielding $ \sin(\alpha_x) \leq (p_e^z)^{1/2} $. 
Using $p_e^x \le 2 p_e$ and 
$p_e^z \le 2 p_e$, we obtain
$ \sin(\alpha_z) \leq (2 p_e)^{1/2} $ and  
$ \sin(\alpha_x) \leq (2 p_e)^{1/2} $. 
In the sequel we will simply drop the indices $x$ and
$z$, 
taking as a convention that we are dealing with the actual basis that
Alice and Bob agreed upon (and which become known to Eve only after
she retransmitted the particle towards Bob).

\paragraph{The state in Eve's hands}

We now look at the $n'$ remaining 
qubits after Alice and Bob discard those
bits where the bases did not agree. Some bits are used to 
verify that $p_{test} \le p_{allowed}$,
to be left with $n$-bit string ${\bf x}$.
{}From the previous paragraph, 
we know that after retransmitting the i-th bit (namely, $x_i$) to 
Bob,
the purification of Eve's state 
is 
$ \ket{ \Psi_{x_i} } =
\cos(\alpha_i)\ \ket{0}_i + (-1)^{x_i} \sin(\alpha_i)\ \ket{1}_i
$,
where $x_i$ is either 0 or 1 ($1 \leq i \leq n$) according to the bit which 
Alice sent to Bob, and $\ket{b}_i$ would be $\ket{b_{{\cal H}_i}}$ in the
notations of the previous paragraph. 
Moreover
$ \sin(\alpha_i) \le (2 p_i)^{1/2}$,
where $p_i$ is Bob's probability of error on the i-th bit
(averaged over the four possible input states),
which is completely determined by Eve's transformation.
The global state of Eve's probes is, thanks to the
properties of the trace, a partial trace of $\ket{\Psi_{\bf x}}$, the tensor 
product of the $\ket{{\Psi_{x_i}}}$. 

To expand $\ket{\Psi_{\bf x}}$ we first need some notations.
Boldface letters like ${\bf j}$, ${\bf x}$ are used to denote
strings in $\{0,1\}^n$ that are interpreted as $n$-vectors on the binary
field. Boldface letters are also used in kets, with the following
understanding: 
if ${\bf j} = j_1 \ldots j_n$ is concatenation of $n$ bits then 
$
\ket{\bf j} =  \ket{j_1}_1 \ldots \ket{j_n}_n
$
where $| b \rangle_i $ are the basis vectors of the purifications
of Eve's i'th qubit.
The state 
$ \ket{\Psi_{\bf x}} = \bigotimes_{i=1}^n 
 [ (\cos(\alpha_i)\ \ket{0}_i + (-1)^{x_i} \sin(\alpha_i)\ \ket{1}_i) ]
$
can be written as 
$ 
\ket{\Psi_{\bf x}} = 
\sum_{{\bf j} \in \{0,1\}^n} d_{\bf j} (-1)^{{\bf x} \cdot {\bf j}}\ \ket{\bf j}
$,
where $d_{\bf j} = d_{j_1} \ldots d_{j_n}$ with $d_{j_{i}}=\cos \alpha_i$ if
${j}_i = 0$ and $d_{j_{i}}=\sin \alpha_i$ if 
${j}_i = 1$,
and where ${\bf x} \cdot {\bf j}$ is by definition 
$
{\bf x} \cdot {\bf j} = \sum_{i=0}^n x_ij_i \ mod \ 2
\ .
$
For instance, $| \Psi_{01} \rangle = \cos \alpha_1 \cos \alpha_2 |00 \rangle
- \cos \alpha_1 \sin \alpha_2 |01\rangle + \sin \alpha_1 \cos \alpha_2 |10 \rangle - \sin \alpha_1 \sin \alpha_2 |11 \rangle$.
We let ${\bf j} \oplus {\bf k}$ be the string obtained by adding
${\bf j}$ and ${\bf k}$ bit by bit with the understanding that 
$1 \oplus 1 = 0$. Then
[using $(-1)^{\bf x \cdot j} (-1)^{\bf x \cdot k} = (-1)^{\bf x\cdot 
{(j \oplus k)}}$], the lift-up of Eve's density matrix is
\begin{equation}
\widetilde{\rho_{\bf x}} = 
\ket{\Psi_{\bf x}}\bra{\Psi_{\bf x}} = 
  \sum_{{\bf j, k} \in \{0, 1\}^n} 
    {{d_{\bf j}}}{{d_{\bf k}}} 
       (-1)^{\bf x \cdot (j \oplus k)}\ \ket{\bf j}\bra{\bf k}
\ ,
\label{psi_x}
\end{equation}
for any string ${\bf x}$ sent by Alice. 

\paragraph{The parity bit}

In order to encode one key-bit $b$ (0 or 1) using a substring of
the $n$ bits she sent,
Alice proceeds as follows:
she chooses some string ${\bf v} \in \{0, 1\}^n$ to define the relevant 
(privacy amplification) substring, and announces it to Bob;
Bob understands that the key-bit sent is $ b ={\bf x \cdot v}$, 
and 
can calculate the final bit $b$. Eve now knows 
${\bf v}$ (but not ${\bf x}$) and has to guess $b = {\bf x \cdot v}$.
Only strings ${\bf x}$ such that ${\bf x \cdot v} = b$
shall contribute to 
$\widetilde{\rho^{{}_{\bf v}}_{b}} \equiv  2^{-n+1}  
\sum_{\{{\bf x} | {\bf x \cdot v} = b\}} 
\ket{\Psi_{\bf x}}\bra{\Psi_{\bf x}} $. 
To learn $b$ Eve needs to distinguish
between the two density matrices (in her hands)
$\rho_b^{\bf v}$
for which 
$\widetilde{\rho_b^{{}_{\bf v}}}$ are lift-ups.
For convenience let us define
$\Delta^{\bf v} \equiv 
\widetilde{\rho^{{}_{\bf v}}_0} - \widetilde{\rho^{{}_{\bf v}}_1} $,
and in the following we evaluate the trace-norm of $\Delta^{\bf v}$. 

Using  $(-1)^b= (-1)^{{\bf x} \cdot {\bf v}}$
and~(\ref{psi_x}) we get 
$
\Delta^{\bf v}  = 
(-1)^0 \widetilde{\rho^{{}_{\bf v}}_0} + 
(-1)^1 \widetilde{\rho^{{}_{\bf v}}_1} =
2^{-n + 1} \sum_{\bf j, k}
{{d_{\bf j}}} {{d_{\bf k}}}
\sum_{\bf x}
(-1)^{\bf x \cdot (j \oplus k \oplus v) }\ \ket{\bf j}\bra{\bf
k} \ .
$
We now simplify the preceding sum using a technique similar to
the one of~\cite{BM97a}.
If ${\bf j  \oplus  k  \oplus  v \neq 0}$, there is some string ${\bf y}$ such that
${\bf(j  \oplus  k  \oplus  v) \cdot y} = 1$. 
If we let
${\bf x'} = {\bf x \oplus y}$  then
$(-1)^{\bf x' \cdot (j  \oplus  k  \oplus  v)} + (-1)^{\bf  x \cdot (j 
\oplus k  \oplus  v)} = 0$ 
and since 
${\bf x \neq x'}$ 
(because ${\bf y \neq 0}$)
{\em all} the coefficients of $\ket{\bf j}\bra{\bf k}$ cancel in pairs.
If ${\bf j  \oplus  k  \oplus  v = 0}$, then
$(-1)^{\bf x \cdot (j  \oplus  k  \oplus  v)} = 1$ for all ${\bf x}$ and since
there are $2^n$ such strings, we get
\begin{equation}
\Delta^{\bf v} 
= 2 \sum_{\bf i \oplus j = v}d_{\bf i}d_{\bf j}\ \ket{\bf i}\bra{\bf j}
= 2 \sum_{{\bf j} \in \{0,1\}^n}
{{d_{\bf j}}} {{d_{\bf j \oplus v}}} \ \ket{\bf j}\bra{\bf j \oplus v}.
\label{dif-noECC1}\end{equation}
If ${\bf i \oplus j = v}$ then clearly ${\bf j \oplus i = v}$.
Therefore, $\Delta^{\bf v}$ is a sum of $2^{n-1}$ Hermitian matrices 
$ d_{\bf i}d_{\bf j}\ket{\bf i}\bra{\bf j}+
 d_{\bf j}d_{\bf i}\ket{\bf j}\bra{\bf i} = 
d_{\bf i}d_{\bf j} [\ket{\bf i}\bra{\bf j}+\ket{\bf j}\bra{\bf i}]$.
For each of them the Trace-norm is
$2 d_{\bf i}d_{\bf j} = d_{\bf i}d_{\bf j} + d_{\bf j}d_{\bf i}$.
Using this result and
the triangle inequality (which is satisfied by any norm) we obtain 
\begin{equation}
Tr | \Delta^{\bf v} | \le 
 2 \sum_{\bf i \oplus j = v}d_{\bf i}d_{\bf j}
 = 2 \sum_{\bf j}
     {{d_{\bf j}}} {{d_{\bf j \oplus v}}}
\ . \label{dif-noECC2}
\end{equation}
If $ v_i$, the $i^{\rm th}$ bit of ${\bf v}$ equals 1, then the product 
of the $i^{\rm th}$ factor of $d_{\bf j}$ by the $i^{\rm th}$ 
bit of $d_{\bf v\oplus j}$
is $\cos \alpha_i \sin \alpha_i$, since  either 
[$d_{j_i}= \cos \alpha_i$ and 
$d_{j_i \oplus v_i}= \sin \alpha_i$]
or alternatively 
[$d_{j_i}= \sin \alpha_i$ and 
$d_{j_i \oplus v_i}= \cos \alpha_i$], since
$j_i \oplus 1 = {\rm not}(j_i)$.
The contribution of such terms is 
$(\sin 2 \alpha_i) $ since the sum is over all ${\bf j}$ so the term
$ {d_{\bf j}} {d_{\bf j \oplus v}}$ contributes twice.
If $ v_i$, the $i^{\rm th}$ bit of ${\bf v}$ equals 0, then the product 
of the $i^{\rm th}$ factor of $d_{\bf j}$ by the $i^{\rm th}$ 
bit of $d_{\bf v\oplus j}$
is either $\cos^2 \alpha_i$ or $ \sin^2 \alpha_i$.
When summing over all ${\bf j}$, such terms sum up to yield $1$.

As result the sum reduces to
$ 
\sum_{\bf j}
{{d_{\bf j}}} {{d_{\bf j \oplus v}}} = 
\prod_{v_i=1} \sin(2\alpha_i) \prod_{v_i=0} 1 
= \prod_{v_i=1} \sin(2\alpha_i)  
$.
If we look at
${\bf v}$ as the characteristic function of a set also denoted ${\bf v}$,
one can write $i \in {\bf v}$ instead of ${v_i = 1}$ and
\begin{equation}
 Tr | \Delta^{\bf v} | \le 2 \prod_{i \in {\bf v}} \sin(2\alpha_i)
\ . \label{dif-noECC3}
\end{equation}
Due to (\ref{Jeroen}) and (\ref{TrMI}) we get
$SD(\rho_0^{\bf v},\rho_1^{\bf v})\le 
SD(\widetilde{\rho^{{}_{\bf v}}_0},\widetilde{\rho^{{}_{\bf v}}_1})\le
Tr | \Delta^{\bf v} | \le 2 \prod_{i \in {\bf v}} \sin(2\alpha_i) $,
if the error correction data is unknown to Eve.
 
\paragraph{Error correction}

For error correction, a number of linear constraints are imposed on the 
bits of ${\bf x}$. More precisely, Alice chooses a  system 
${\bf E}  = \{ {\bf v}_1 \cdot {\bf x} = b_1, 
{\bf v}_2 \cdot {\bf x} = b_2, \ldots,
{\bf v}_r \cdot {\bf x} = b_r \}$  
of $r$ linear equations 
such that the $r+1$ strings
${\bf v}$, ${\bf v}_1$, \ldots, ${\bf v}_r$ are linearly
independent and each $b_i$ is either 0 or 1. We write ${\bf E(x)}$
to mean that ${\bf x}$ satisfies the system (that is, ${\bf x}$
is a code word).

We can now define $\Delta^{\bf E,v}$ as in the previous section
with $\widetilde{\rho^{{}_{\bf E,v}}_b}$ 
an equal mixture of the states
$\ket{\Psi_{\bf x}}\bra{\Psi_{\bf x}}$ 
such that ${\bf x}$ satisfies the system
$\{{\bf x \cdot v} = b\} \cup {\bf E}$. 
This system has 
$2^{n - r - 1}$ solutions and so
$
\Delta^{\bf E,v} = 2^{-n + r + 1} \sum_{\bf j, k}
{{d_{\bf j}}} {{d_{\bf k}}}
\sum_{\bf E(x)}
(-1)^{\bf x \cdot (j \oplus k \oplus v) }\ \ket{\bf j}\bra{\bf
k}
$.
As in the previous section, the expression for the Trace-norm
can be simplified. 
For any ${\bf s} \in \{ 0, 1 \}^r$ 
let ${\bf v_s}$ denote the element
$\sum_{i=1}^r  s_i {\bf v}_i$. If ${\bf j  \oplus  k  \oplus  v } = {\bf v_s}$
and ${\bf x}$ is a solution of ${\bf E}$, then the exponent
${\bf x \cdot (j \oplus k \oplus v)}$ in the above expression for 
$\Delta^{\bf E,v} $ reduces to
$(\sum_{i=1}^r  s_i {\bf v}_i ) \cdot {\bf x}$ = 
$\sum_{i=1}^r  s_i\ {\bf v}_i \cdot {\bf x}$ = 
$\sum_{i=1}^r s_i b_i$ = ${\bf s \cdot b}$.
This value is independent of
${\bf x}$ and so the coefficient of 
$\ket{\bf j}\bra{\bf k}$ = $\ket{\bf j}\bra{{\bf k  \oplus  v  \oplus  v_s}}$ 
is
$2 d_{\bf j}d_{{\bf k}  \oplus  {\bf v}  \oplus  
{\bf v_s}}(-1)^{\bf s \cdot b}$ 
where ${\bf b}$ is the string $(b_i)_{1 \leq i \leq r}$ of the parity bits
in the equations of ${\bf E}$.
If ${\bf j  \oplus  k  \oplus  v}$
is not in the span of $\{{\bf v}_1$, \ldots, ${\bf v}_r\}$ 
then there is a solution
${\bf y}$ to the system 
 $\{{\bf (j  \oplus k  \oplus  v) \cdot y} = 1$, ${\bf v}_1 \cdot
{\bf y} = 0, \ldots, {\bf v}_r \cdot {\bf y} = 0\}$. For any ${\bf x}$ solution
of ${\bf E}$, let ${\bf x'}$ denote ${\bf x  \oplus  y}$. Clearly ${\bf x'}$ is
also a solution of ${\bf E}$ and 
$(-1)^{\bf x \cdot (j \oplus k \oplus v) }
+ (-1)^{\bf x' \cdot (j \oplus k \oplus v) } = 0$,
and consequently the coefficient of 
$\ket{\bf j}\bra{\bf k}$ is 0.
Therefore 
\begin{equation}
\Delta^{\bf E,v}  = 2 
\sum_{{\bf j}, {\bf s}}
{{d_{\bf j}}} d_{{\bf j}  \oplus  {\bf v}  \oplus  {\bf v}_s}
(-1)^{{\bf s \cdot b}}\ \ket{\bf j}\bra{\bf
j  \oplus  v  \oplus  v_s}
\label{dif-ECC}
\ , 
\end{equation}
generalizing Eq.~(\ref{dif-noECC1}) to contain the error correction data.
Consequently
$ 
\Delta^{\bf E,v} =  
\sum_{{\bf s} \in \{0,1\}^r} (-1)^{\bf s \cdot b} (
  \widetilde{\rho_0^{{}_{{\bf v}  \oplus  {\bf v_s}}}} - 
\widetilde{\rho_1^{{}_{{\bf v}  \oplus  {\bf v_s}}}} )
$. 
As before we define  
$ \Delta^{{\bf v}  \oplus  {\bf v_s}} $ for the terms in the parenthesis,
and these terms are given by Eq.~(\ref{dif-noECC1}) [and their Trace-norm 
is given by Eqs.~(\ref{dif-noECC2}) and~(\ref{dif-noECC3})]
once ${\bf v}$ there is replaced by ${\bf v \oplus v_s}$. 
This gives 
$ \Delta^{\bf E,v}  =  
\sum_{{\bf s} \in \{0,1\}^r}
(-1)^{{\bf s \cdot b}} (\Delta^{\bf v \oplus v_s})$, and due
to the triangle inequality 
$ 
Tr | \Delta^{\bf E,v} | \leq
\sum_{{\bf s} \in \{0,1\}^r} 
  Tr | \Delta^{{\bf v}  \oplus  {\bf v_s}} |
$.
Using the set notation and Eq.~(\ref{dif-noECC3}) we finally get
\begin{equation}
Tr | \Delta^{\bf E,v} | \leq
  2 \sum_{{\bf s} \in \{0,1\}^r} \prod_{i \in ({\bf v} \oplus {\bf v_s})} 
   \sin(2\alpha_i)
\ .
\end{equation}
Due to (\ref{Jeroen}) and (\ref{TrMI}),  we get 
$SD(\rho_0^{\bf E,v},\rho_1^{\bf E,v})\le 
SD(\widetilde{\rho^{{}_{\bf E,v}}_0},\widetilde{\rho^{{}_{\bf E,v}}_1})\le
Tr | \Delta^{\bf E,v} | \leq
  2 \sum_{{\bf s} \in \{0,1\}^r} \prod_{i \in ({\bf v} \oplus {\bf v_s})} 
   \sin(2\alpha_i)  $
when the error correction data is known to Eve.
Using $\sin (2\alpha_i) \le 2 \sin \alpha_i \le (8 p_i)^{1/2}$ we finally get
$SD(\rho_0^{\bf E,v},\rho_1^{\bf E,v})\le 
  2 \sum_{{\bf s} \in \{0,1\}^r}  
  [\prod_{i \in ({\bf v} \oplus {\bf v_s})} 
   (8 p_i )]^{1/2} $.

Let the ``Hamming weight'' $\hat n_{\bf s}$ 
(for each ${\bf s}$)
be the number of one's in ${\bf v \oplus v_s}$ 
[the number of factors in the product
$\prod_{i \in ({\bf v} \oplus {\bf v_s})}$]. 
Also let $p_{\bf s} = [ \sum_{i \in ({\bf v} \oplus {\bf v_s})} p_i ]/ 
\hat n_{\bf s}$ be 
the average error in any 
relevant subset ${\bf s}$.
The geometrical mean of the $p_i$ contributing to $p_{\bf s}$ is always
less than their arithmetical mean so 
$ [ \prod_{i \in ({\bf v} \oplus {\bf v_s})} 
   (8 p_i ) ]^{1/2}  \le 
[ 8 p_{\bf s}]^{\hat n_{\bf s} / 2}$, and thus 
$SD(\rho_0^{\bf E,v},\rho_1^{\bf E,v})\le 
        2 \sum_{{\bf s} \in \{0,1\}^r}  [ 8 p_{\bf s}]^{\hat n_{\bf s} / 2}$.

Given that the test is passed $p_{test} \le p_{allowed}$ 
statistical analysis promise us 
that each of the $p_{\bf s}$ is bounded. 
Combining two laws of large numbers of Hoeffding~\cite{Hoeffding}, 
Theorem 2 (sums of independent random variables)
and its extension in section 6 (sampling from a finite population),
we are promised that $p_{n'}$, the average
$p_i$ of all $n'$ relevant bits satisfies
${\rm Prob} [p_{n'} > p_{test} + 2\delta] \le 2 e^{-2 n_{test} \delta^2}$ 
(since the tested bits are picked at random).
Once $p_{n'}$ is bounded we can bound $p_{\bf s}$ as follows:
let $n'$ be even [throw one bit if needed (before choosing the bits for
the test)], and 
let Alice and Bob use  $n_{test} = n'/2$ bits for the test.
We then have  
$p_{\bf s} \le (n'/\hat{n}_{\bf s}) p_{n'}$.
Thus, Eve's information is generously bounded by 
$2 \sum_{{\bf s} \in \{0,1\}^r} [ (8n'/\hat{n}_{\bf s}) (p_{test} + 
2 \delta)]^{(\hat n_{\bf s} / 2)} $,
except with a probability of $p_{luck} = 2 e^{-2 n_{test} \delta^2}$. 
Recall that $n_{test} = n = n'/2$.
Assuming (generously again) that in such a case of having {\em luck}
Eve's information is maximal (one bit)
her total information is bounded by 
$2 \sum_{{\bf s} \in \{0,1\}^r} [(16n/\hat{n}_{\bf s})(p_{test} + 
2 \delta)]^{(\hat n_{\bf s} / 2)} + 
2 e^{-2 n \delta^2}$, for any $\delta$. 
Let $\alpha n = \hat{n} = \min_{\bf s} \hat n_{\bf s}$.
Then, 
$SD(\rho_0^{\bf E,v},\rho_1^{\bf E,v})\le 
2^{r+1} [(16/\alpha)(p_{test} + 
2 \delta)]^{\alpha n / 2} + 
2 e^{-2 n \delta^2}$.
Entering into coding theory is beyond our aim in this letter 
and is left for the full paper:
for error rates below $2\%$, many codes allow us to choose the parameters
$n$, $r$, $\alpha$ and $\delta$ such that Eve's information is negligible
[e.g., $2^{-100}$ of a bit].

\end{multicols}

\end{document}